\documentclass[aps,prd,superscriptaddress,showkeys,showpacs,twocolumn,longbibliography,nofootinbib]{revtex4-1}
\usepackage{amsmath}
\usepackage{float}
\usepackage{braket}
\usepackage{graphicx}
\usepackage{enumitem}
\usepackage{multirow}
\usepackage[colorlinks=true, pdfstartview=FitV, linkcolor=red, citecolor=blue, urlcolor=blue]{hyperref}
\usepackage{slashed}

\begin{document}
\title{
Heavy quark energy loss in the Bayesian-inference-improved CUJET model
}

\author{Yu Guo}  
\address{Department of Physics, Tsinghua University, Beijing 100084, China.}

\author{Jinfeng Liao} \email{liaoji@iu.edu}
\address{Physics Department and Center for Exploration of Energy and Matter, Indiana University, 2401 N Milo B. Sampson Lane, Bloomington, IN 47408, USA.}

\author{Shuzhe Shi} \email{shuzhe-shi@tsinghua.edu.cn}
\address{Department of Physics, Tsinghua University, Beijing 100084, China.}
\affiliation{State Key Laboratory of Low-Dimensional Quantum Physics, Tsinghua University, Beijing 100084, China.}

\date{\today}

\begin{abstract}
We compute jet quenching observables---the nuclear modification factor $R_{AA}$ and elliptic anisotropy coefficient $v_2$---for open-heavy flavor mesons with high transverse momentum in an improved CUJET model, in which the strong coupling parameter as well as  the temperature-dependent fraction of chromo-magneto monopoles  are extracted through  Bayesian inference using high-$p_T$ {\em light-hadron} $R_{\mathrm{AA}}$ and $v_2$ measurements from RHIC and the LHC. 
We show in this work that such a Bayesian-inference-improved CUJET model  provides a good description of available high-$p_T$ charm- and bottom-hadron observables based on the same posterior parameters.  
The corresponding heavy-quark spatial diffusion parameter, $2\pi T D_s$, is also computed, which is broadly consistent with current phenomenological and lattice-QCD constraints. 
These results support a universal  description for the medium modification of high-$p_T$ light- and heavy-flavor energy loss and indicate an important role played by the chromo-magnetic degrees of freedom near the confinement transition region of quantum chromodynamics (QCD). 
\end{abstract}

\maketitle

\section{Introduction}
\label{sec:introduction}

The quark-gluon plasma (QGP) — a strongly interacting many-body system in the color-deconfined phase — is created in relativistic heavy-ion collisions at RHIC and the LHC.
The formation of the QGP is supported by various independent phenomena observed in experiments, including pronounced collective behavior, suppression of particles with high transverse momentum (also referred to as ``jet quenching''), and modification of quarkonium production,  etc. 
Modern experimental and phenomenological studies of QGP physics have entered the precision era, which enables quantitative constraints on the thermodynamic and transport properties of the QGP. Such information provides crucial insights into  the fundamental question of understanding the effective microscopic degrees of freedom in the QGP near the phase boundary of quantum chromodynamics (QCD), which is closely related to the central challenge of strong interaction physics — the microscopic origin of QCD confinement~\cite{tHooft:1999cgx, Greensite:2011zz, Gross:2022hyw}.

Chromo-magnetic monopoles (CMMs) --- emergent topological configurations of non-Abelian gauge fields that carry chromo-magnetic charge and source long-range magnetic fields --- have long been perceived to play a prominent role in QCD confinement~\cite{tHooft:1974kcl, Polyakov:1974ek}. In the QCD vacuum, the CMM condensate is conjectured to be the underlying mechanism for the ’t Hooft-Mandelstam dual superconductor picture, in which chromo-electric charges are confined into hadrons by the magnetic condensate~\cite{Mandelstam:1974pi, tHooft:1977nqb, tHooft:1981bkw}. 
When the temperature is increased to around the hadron-parton transition temperature $T_c$, the CMM degrees of freedom are  expected to ``evaporate'' into thermal quasiparticles and coexist with the chromo-electric quarks and gluons in the $1\sim2T_c$ region~\cite{Liao:2005pa,Liao:2006ry,Liao:2008jg,Liao:2008dk,Chernodub:2006gu,DAlessandro:2007lae,Bonati:2013bga,Hidaka:2008dr,Hidaka:2009ma,Shuryak:2008eq,Ratti:2008jz,Ramamurti:2017zjn,Fujimoto:2025sxx}. An important question here is to look for potential phenomenological manifestations of such CMMs. 

Jet quenching provides a sensitive probe of the microscopic degrees of freedom in the QGP. Energetic quarks and gluons produced in initial hard scatterings propagate through the evolving QGP and interact with its constituents before fragmenting into final-state hadrons~\cite{Wang:1992qdg, Gyulassy:2000gk}. These interactions induce energy loss of the hard parton through elastic energy transfer and medium-induced gluon radiation processes. 
If the QGP medium indeed includes both electric quasiparticles (like quarks and gluons) and magnetic quasiparticles (the CMMs), then the jet parton interacts differently with them. In particular, the interaction strength between a jet parton and a chromo-magnetic monopole is required by the famous Dirac quantization condition to be always strong: $\alpha_E\,\alpha_M=1$ where $\alpha_E$ and $\alpha_M$ are electric and magnetic couplings respectively~\cite{Dirac:1931kp, Liao:2008jg, Liao:2008dk}. Thus, the jet-medium interaction can be significantly enhanced in the temperature region where the CMMs become abundant. Such an effect can lead to increased jet energy loss in the QGP with measurable consequences.  

Experimentally, the energy loss of energetic particles can be measured as a suppression of high-transverse-momentum $p_T$ hadron production relative to an incoherent superposition of proton–proton collisions. The suppression and its azimuthal dependence are characterized by the nuclear modification factor
\begin{align}
\begin{split}
R_{\mathrm{AA}}(p_T,\phi)
\equiv{}&
\frac{
\mathrm{d}^3N_{\mathrm{AA}}(p_T,\phi)/
p_T\mathrm{d}p_T\,\mathrm{d}\phi\,\mathrm{d}\eta
}{
N_{\mathrm{bin}}
\mathrm{d}^3N_{pp}(p_T,\phi)/
p_T\mathrm{d}p_T\,\mathrm{d}\phi\,\mathrm{d}\eta
}
\\
={}&
R_{\mathrm{AA}}(p_T)
 [
1+2v_2(p_T)\cos(2\phi-2\Psi_2)+\cdots ],
\label{eq:raa_definition}
\end{split}
\end{align}
where $N_{\mathrm{bin}}$ denotes the number of binary nucleon–nucleon collisions and $\Psi_2$ is the second-order event-plane angle. The angle-averaged $R_{\mathrm{AA}}$ primarily constrains the average strength of parton energy loss, whereas the elliptic coefficient $v_2$ is more sensitive to the  path-length and temperature dependence of energy loss. The jet quenching phenomenon has been studied with several different  formalisms that have varied assumptions of medium constituents and/or different treatments of perturbative expansions in the gluon emission process. They include e.g.  \textsc{amy}~\cite{Arnold:2001ba, Arnold:2001ms, Arnold:2002ja}, \textsc{asw}~\cite{Wiedemann:2000za, Salgado:2003gb, Armesto:2003jh, Armesto:2004pt,Armesto:2005iq}, \textsc{dglv}~\cite{Gyulassy:1993hr, Gyulassy:2000er, Djordjevic:2003zk, Djordjevic:2008iz, Buzzatti:2011vt, Xu:2014ica}, \textsc{ads-cft}~\cite{Casalderrey-Solana:2014bpa}, \textsc{lbt}~\cite{He:2015pra, Cao:2016gvr}, higher-twist approaches~\cite{Guo:2000nz, Wang:2001ifa, Majumder:2007ae}, and geometric energy-loss models~\cite{Shuryak:2001me, Drees:2003zh, Betz:2014cza, Noronha-Hostler:2016eow}.

The \textsc{cujet3} model~\cite{Xu:2014tda, Xu:2015bbz, Shi:2018lsf, Shi:2018izg} is a comprehensive framework for describing jet energy loss based on a microscopic description of QGP that quantitatively implements the nonperturbative suppression of quarks and gluon and the inclusion of the CMM degrees of freedom. It has been shown to successfully provide a simultaneous description of the $R_{\mathrm{AA}}$ and $v_2$ observables across different collision systems and beam energies. Recently in Ref.~\cite{Guo:2025tsf}, we performed a Bayesian inference of the temperature-dependent chromo-magnetic fraction in the QGP by utilizing a large set of experimental data on light hadron observables. The Bayesian analysis reveals a substantial chromo-magnetic component in the $1\sim2\,T_c$ region, and the data strongly favor this scenario over a purely chromo-electric medium. 

Heavy quark energy loss offers a complementary category of sensitive probe~\cite{Dong:2019byy, Dong:2019unq}, in which the charm and bottom quarks are produced predominantly in the initial hard processes and subsequently interact with the QGP throughout their dynamical evolution. Their finite masses modify the relative contributions of radiative and collisional energy loss and lead to flavor-dependent quenching patterns~\cite{Djordjevic:2003zk, Wicks:2005gt}. Measurements of charm- and bottom-hadron suppression and azimuthal anisotropy can consequently provide an independent test of the Bayesian-inference-improved CUJET3 model developed in ~\cite{Guo:2025tsf}. 
In this work, we apply the posterior distribution of the \textsc{cujet3} parameters from the Bayesian analysis to calculate $R_{\mathrm{AA}}$ and $v_2$ observables in the heavy flavor sector and to estimate the  heavy-quark spatial diffusion coefficient $D_s$. Such analyses together will provide validation on whether the same microscopic model of QGP can  account for energy loss and transport properties for both light-flavor and heavy flavor sectors.  
The rest of the paper is organized as follows. In Sec.~\ref{sec:bayesian} we briefly review the Bayesian analysis and provide additional details supporting the results shown in Ref.~\cite{Guo:2025tsf}. Results for  heavy flavor jet-quenching observables and diffusion coefficient are shown and discussed in Sec.~\ref{sec:heavy_flavor}. Finally, a summary is  provided in Sec.~\ref{sec:summary}.

\section{Bayesian constraints on the QGP composition in the CUJET framework}
\label{sec:bayesian}
For completeness, we begin by reviewing the Bayesian analysis (BA) of the QGP composition based on the light flavor observables~\cite{Guo:2025tsf}. In what follows, we will briefly introduce the \textsc{cujet} model and the BA framework. Complementary details of the BA analysis beyond those reported in \cite{Guo:2025tsf} will also be provided.

\subsection{Chromo-magneto monopoles in CUJET}
\label{sec:bayesian:cujet}

The \textsc{cujet} model describes the energy loss of energetic quarks and gluons (jets) through a QCD medium that evolves in space and time. A jet loses energy via either elastic collisions with the medium particle or emitting gluons in inelastic processes. In \textsc{cujet}, the elastic contribution is evaluated using the Thoma--Gyulassy formalism with the Peign\'e--Peshier running-coupling prescription~\cite{Thoma:1990fm, Bjorken:1982tu, Peigne:2008nd}, while the radiative contribution is calculated within the \textsc{dglv} opacity-expansion framework~\cite{Gyulassy:1999zd, Gyulassy:2000er, Djordjevic:2003zk, Djordjevic:2008iz}. While details of the model can be found in Refs.~\cite{Xu:2014ica, Xu:2014tda, Xu:2015bbz, Shi:2018lsf, Shi:2018izg, Guo:2025tsf}, here we highlight the components that are relevant in the BA. First is the running coupling, which takes the infrared-saturated non-perturbative form,
\begin{equation}
    \alpha_s(Q^2)
=
    \frac{\alpha_C}
    {1+\dfrac{9\alpha_C}{4\pi}
    \ln\!\left(Q^2/\Lambda^2\right)},
\qquad
    \Lambda=200~\mathrm{MeV},
\label{eq:running_coupling}
\end{equation}
where the saturation parameter $\alpha_C$ needs to be constrained in the BA. The running coupling controls the strength of the elastic collisions and gluon emission. The probability density of emitting a gluon by a jet parton of energy $E$ with energy fraction $x_E$ is given by
\begin{align}
\begin{split}
&x_E \frac{dN^{n=1}_g}{dx_E} 
\\=\,&
    \frac{3C_R}{\pi^2} \int d\tau\,\Gamma \int d^2 \mathbf{k}_\perp\, \alpha_s\Big(\frac{\mathbf{k}_\perp^2}{x_+(1-x_+)}\Big)
\\& 
    \int d^2 \mathbf{q}_\perp\left(
    (3\rho_g + \frac{4}{3}\rho_q)
    \frac{\mathrm{d}\sigma_{E}}{\mathrm{d}\mathbf{q}_\perp^2}
+
    3\rho_m\frac{\mathrm{d}\sigma_{M}}{\mathrm{d}\mathbf{q}_\perp^2}
    \right)
\\&
    \left(
    \frac{\mathbf{k}_\perp - \mathbf{q}_\perp}{(\mathbf{k}_\perp - \mathbf{q}_\perp)^2 + \chi^2} \cdot
    \Big( 
    \frac{ \mathbf{k}_\perp - \mathbf{q}_\perp }{(\mathbf{k}_\perp - \mathbf{q}_\perp)^2 + \chi^2} 
    -
    \frac{\mathbf{k}_\perp}{\mathbf{k}_\perp^2 + \chi^2} 
    \Big) 
    \right) 
\\&
    \left(
    1 - \cos\Big( \frac{ (\mathbf{k}_\perp - \mathbf{q}_\perp)^2 + \chi^2 }{2x_+E}\tau \Big) 
    \right)
    \;\left| \frac{dx_+}{dx_E} \right|
\,.
\end{split}\label{eq:dNdx}
\end{align}
where $C_R$ is the color factor, $\Gamma$ the fluid Lorentz factor, $x_+$ the fractional plus momentum.

The existence of CMMs modifies the radiative energy loss in two ways. First, the number densities of chromo-electric components, quarks and gluons, get reduced by a temperature dependent factor $\chi_T(T)$,
\begin{align}
    \rho_g(T)
=
\chi_T(T)
\frac{16\,\rho(T)}{16+9N_f},
\quad
    \rho_q(T)
=
\chi_T(T)
\frac{9N_f\,\rho(T)}{16+9N_f},
\label{eq:densities}
\end{align}
and the remainder of the medium are CMMs, $\rho_m(T)=(1-\chi_T(T))\rho(T)$. Here, $\rho(T)$ is the total quasiparticle density, and $N_f$ is the effective number of quark flavors.

Second, the CMMs scatter off the jets and also modifies the effective screening mass ($\mu$) and scattering with chromo-electric components,
\begin{align}
\frac{\mathrm{d}\sigma_E}
{\mathrm{d}\mathbf{q}_\perp^2}
=
    \frac{f_E^2\alpha_s^2(\mathbf{q}_\perp^2)}
    {(\mathbf{q}_\perp^2+f_E^2\mu^2)\mathbf{q}_\perp^2},
\quad
\frac{\mathrm{d}\sigma_M}
{\mathrm{d}\mathbf{q}_\perp^2}
=
    \frac{f_M^2}
    {(\mathbf{q}_\perp^2+f_M^2\mu^2)\mathbf{q}_\perp^2}.
\label{eq:sigma}
\end{align}
The electric and magnetic screening coefficients are
\begin{equation}
f_E(T)=\sqrt{\chi_T(T)},
\qquad
f_M(T)=c_M\sqrt{4\pi\alpha_s(\mu^2)},
\label{eq:screening_coefficients}
\end{equation}
with $c_M$ being another Bayesian parameter. 
Modification of the electric screening further propagates to the parton mass, $\chi^2=M^2x_+^2+\frac{1-x_+}{2}f_E^2\mu^2$.

Apparently, the temperature dependent chromo-electric fraction $\chi_T(T)\in[0,1]$ controls the medium composition, and $1-\chi_T(T)$ gives the chromo-magnetic fraction. Setting $\chi_T(T)=1$ gives $\rho_m(T)=0$ and therefore switches off the CMM component, providing the no-CMM baseline. In our BA~\cite{Guo:2025tsf}, we discretize the $\chi_T(T)$ function on temperature grids
\begin{equation}
T_i\in
\{0.16,0.20,\ldots,0.52\}~\mathrm{GeV},
\label{eq:temperature_nodes}
\end{equation}
and $\chi_T(T)$ is linearly interpolated between adjacent temperatures. 
The model parameters form a twelve-dimensional vector
\begin{equation}
\boldsymbol{\theta}
=
\left\{
\chi_T(T_1),\ldots,\chi_T(T_{10}),
\alpha_C,c_M
\right\}.
\label{eq:parameter_vector}
\end{equation}

We assign uniform priors
\begin{align}
0.3 &\leq \alpha_C \leq1.34,
\nonumber\\
0.19 &\leq c_M \leq 0.56,
\nonumber\\
0 &\leq \chi_T(T_i) \leq 1,
\label{eq:priors}
\end{align}
and consider two prior structures for $\chi_T(T)$. In the \emph{unconstrained} scenario, the ten nodal values are sampled independently, allowing the data to determine the temperature dependence within the resolution of the chosen parameterization. In the \emph{monotonic} scenario, we impose a progressive liberation of chromo-electric degrees of freedom as temperature increases~\cite{Liao:2005pa, Liao:2008jg, Hidaka:2008dr, Hidaka:2009ma},
\begin{equation}
\chi_T(T_1) \leq
\chi_T(T_2) \leq
\cdots \leq \chi_T(T_{10}).
\label{eq:monotonic_prior}
\end{equation}

\subsection{Posterior distribution}
\label{sec:parameter_posteriors}
In~\cite{Guo:2025tsf}, we perform BA with the calibration dataset consisting of high-$p_T$ light-hadron $R_{\rm AA}$ and $v_2$ measurements~\cite{Adare:2012wg, PHENIX:2013yhu, ATLAS:2015qmb, ATLAS:2011ah, Khachatryan:2016odn, Sirunyan:2017pan} in central and semi-central Au--Au collisions at $\sqrt{s_{NN}}=200~\mathrm{GeV}$ and Pb--Pb collisions at $\sqrt{s_{NN}}=2.76$ and $5.02~\mathrm{TeV}$.
Given the experimental dataset $\mathcal D$, Bayes' theorem gives the likelihood distribution of a parameter set $\boldsymbol{\theta}$,
\begin{equation}
p(\boldsymbol{\theta} | \mathcal D)
=
\frac{ p(\mathcal D | \boldsymbol{\theta})\, p(\boldsymbol{\theta})}
{p(\mathcal D)},
\label{eq:bayes}
\end{equation}
where $p(\boldsymbol{\theta})$ is the prior, $p(\mathcal D)$ is the Bayesian evidence, and $p(\mathcal D | \boldsymbol{\theta})$ is the likelihood modeled as a multivariate Gaussian,
\begin{equation}
    \ln p(\mathcal D | \boldsymbol{\theta})
=
    -\frac{1}{2}
    \left(\boldsymbol{\Delta}^{\mathrm T} \cdot \boldsymbol{\Sigma}^{-1} \cdot
    \boldsymbol{\Delta}
    +\ln\det\boldsymbol{\Sigma}
    +N\ln(2\pi) \right).
\label{eq:likelihood}
\end{equation}
Here, $\boldsymbol{\Sigma}$ is the total covariance matrix, and
\begin{equation}
    \boldsymbol{\Delta}
=
    \boldsymbol{y}_{\rm exp}
    - \boldsymbol{y}_{\rm model}(\boldsymbol{\theta}),
\end{equation}
measures the difference between the experimental data and the model prediction. Further discussions of $\boldsymbol{y}_{\rm model}(\boldsymbol{\theta})$ and $\boldsymbol{\Sigma}$ will be given in what follows.

Overall, parameters $\boldsymbol{\theta}$ are sampled according to the posterior distribution $p(\boldsymbol{\theta} | \mathcal D)$, performed with the No-U-Turn Sampler~\cite{hoffman2014nuts}, an adaptive Hamiltonian Monte Carlo algorithm implemented in \textsc{pyro}~\cite{bingham2019pyro}. Selected high-posterior-probability parameter sets are also reevaluated with the full \textsc{cujet} model, confirming that the inferred posterior is not driven by emulator interpolation artifacts.

\begin{figure}[!ht]
\centering
\includegraphics[width=0.3\textwidth]{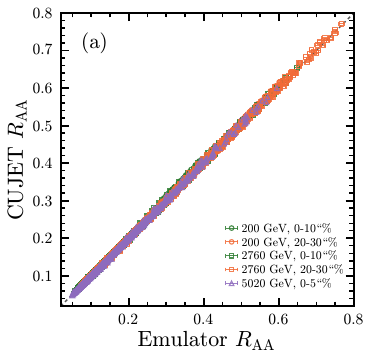}
\includegraphics[width=0.3\textwidth]{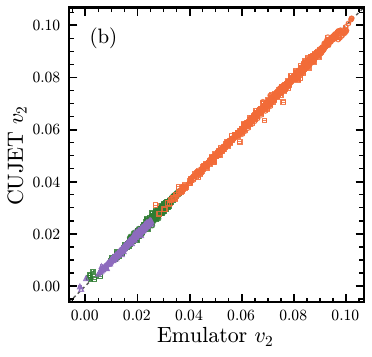}
\caption{
Validation of the Gaussian-process emulator on an independent validation set. Emulator predictions are compared with full \textsc{cujet} calculations for $R_{\rm AA}$ and $v_2$. The diagonal line represents perfect agreement.
}
\label{fig:gpe_validation}
\end{figure}
\begin{figure*}[!ht]
    \centering
    \includegraphics[width=.92\linewidth]{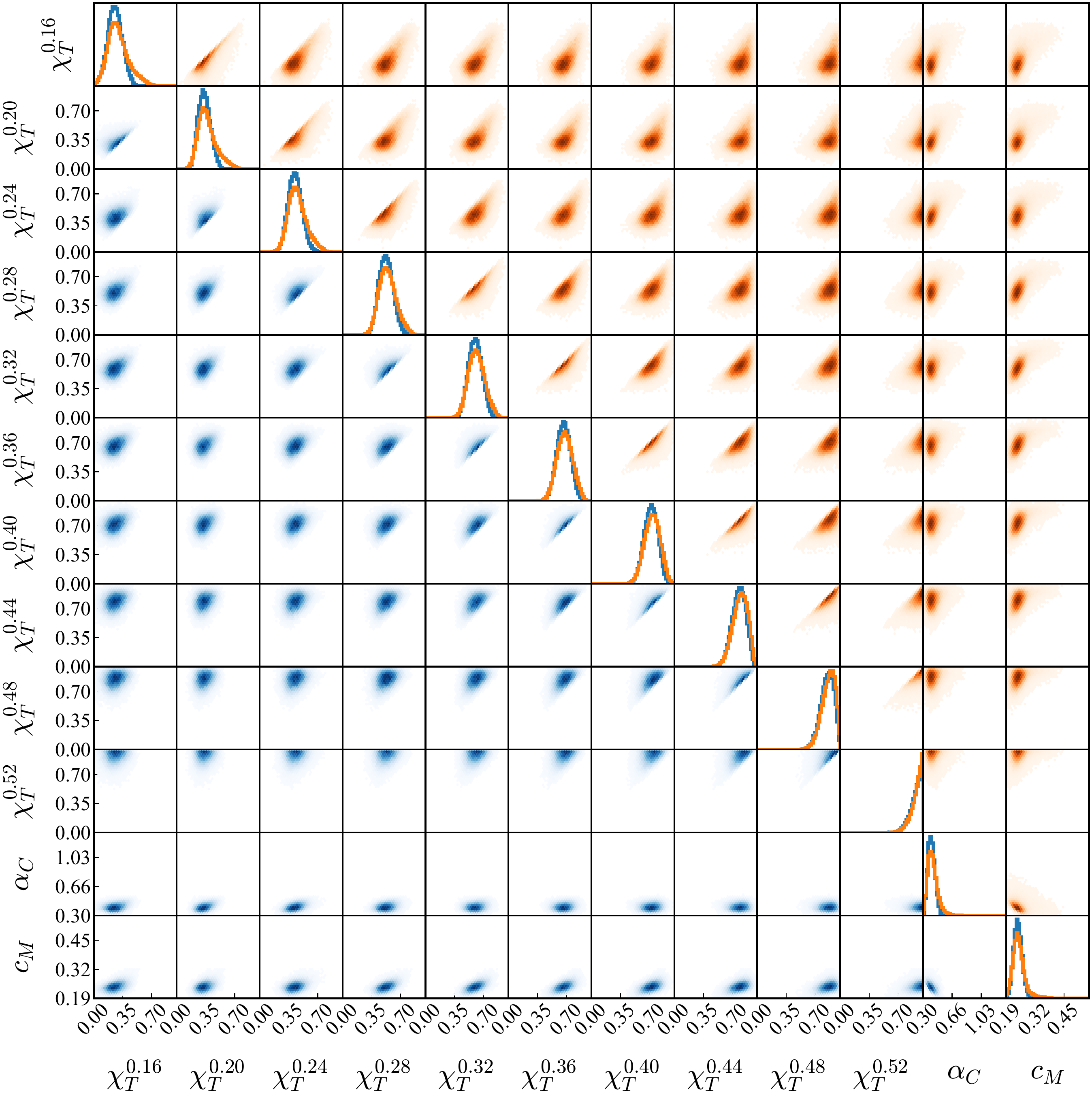}
    \caption{Posterior distributions with correlated experimental uncertainties assuming a relative normalization uncertainty of $\tau = 0.05$. The blue distributions correspond to the baseline analysis with purely diagonal uncorrelated uncertainties, while the orange distributions include the correlated covariance term.}
    \label{fig:CoV_05}
\end{figure*}
\begin{figure*}[!ht]
    \centering
    \includegraphics[width=.92\linewidth]{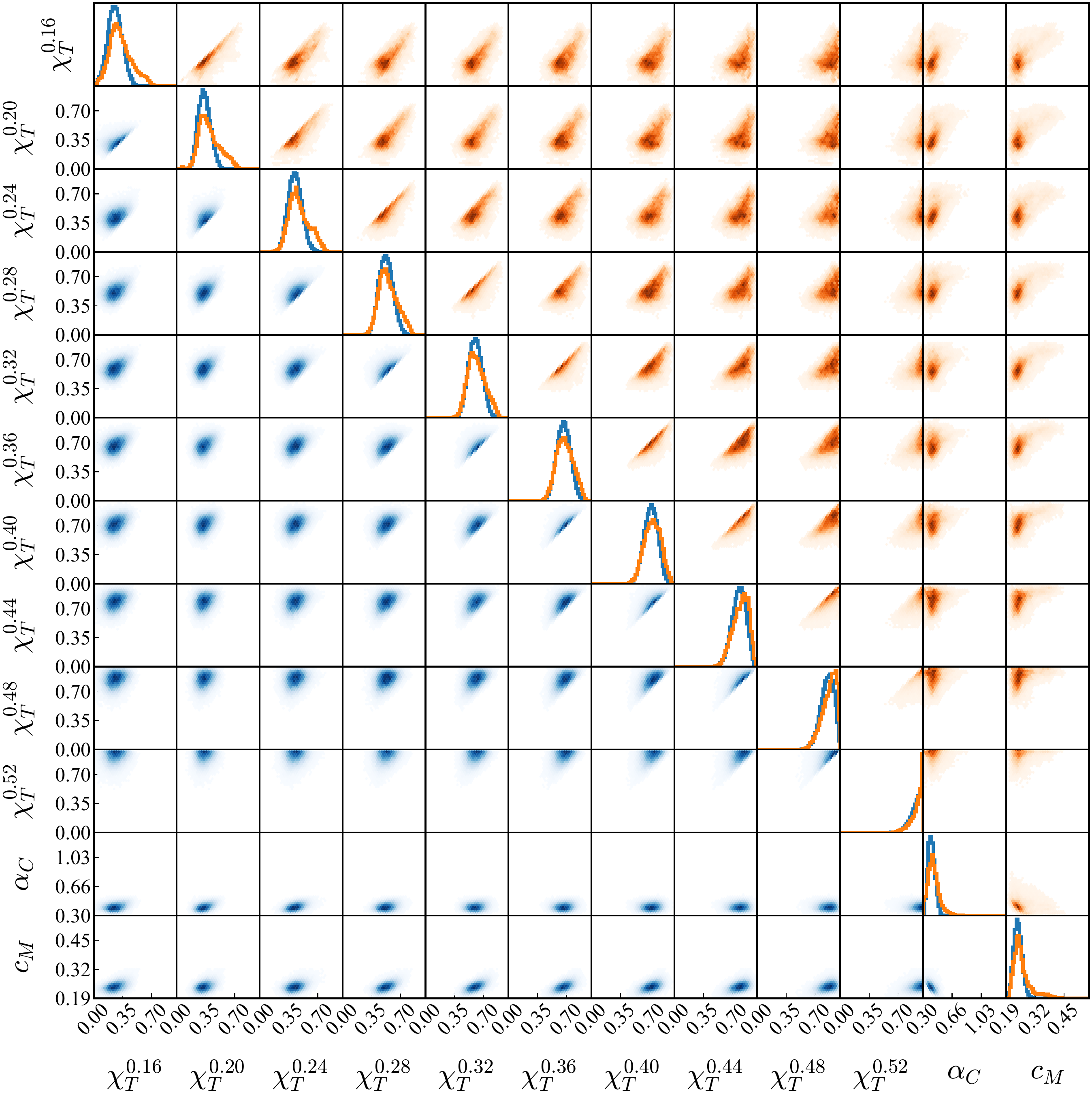}
    \caption{Same as Fig.~\ref{fig:CoV_05}, but with a larger correlation strength $\tau = 0.10$. The posterior distributions become broader, while the inferred temperature dependence and near-$T_c$ enhancement remain robust.}
    \label{fig:CoV_10}
\end{figure*}
\begin{figure*}[!ht]
\centering
\includegraphics[width=.92\linewidth]{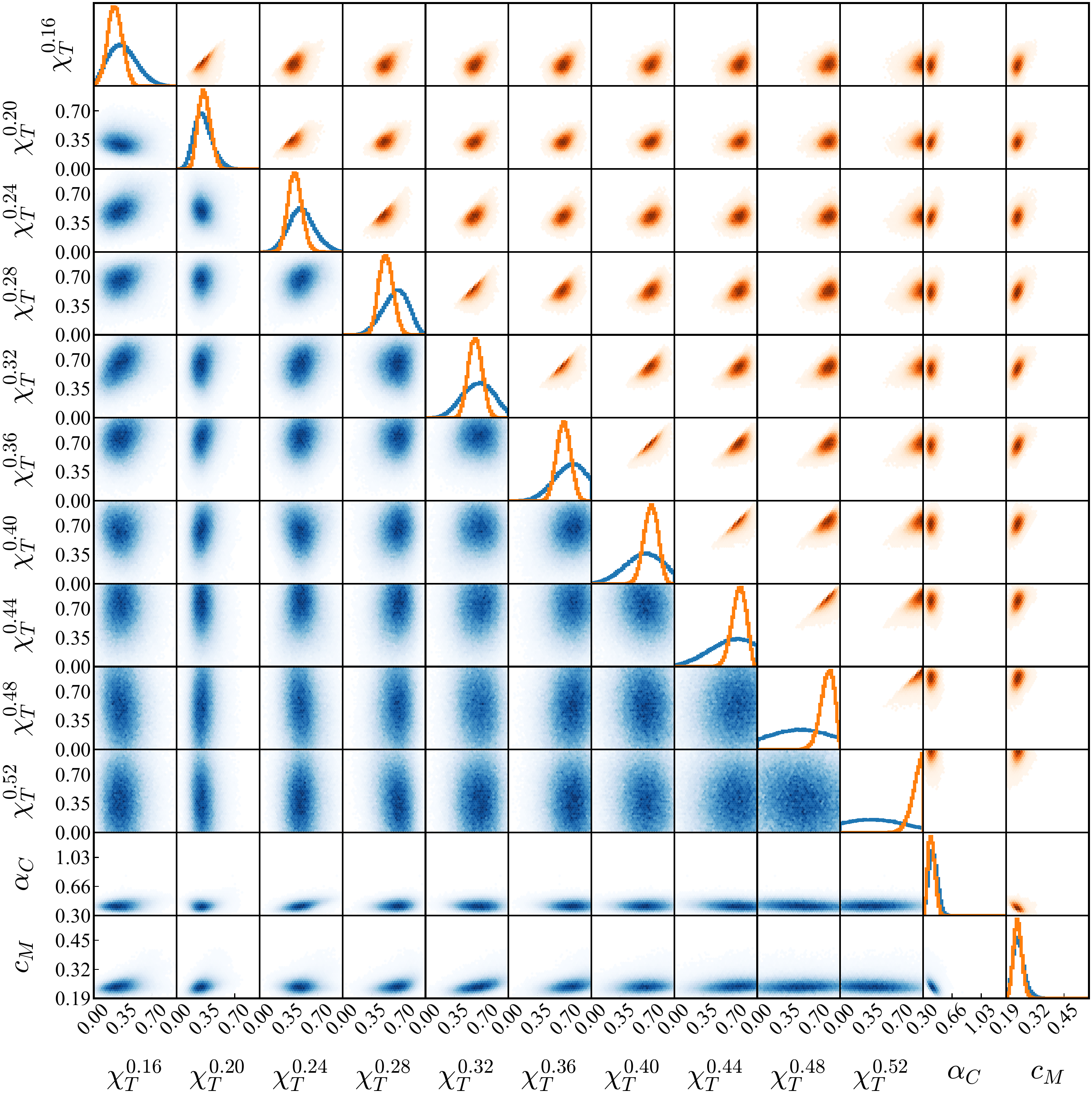}
\caption{
Marginal and pairwise posterior distributions of the model parameters.
Blue distributions in the lower triangle correspond to the
unconstrained scenario, while orange distributions in the upper
triangle correspond to the monotonic scenario.
}
\label{fig:posteriors}
\end{figure*}

\vspace{3mm}
\emph{Gaussian-process emulator of $\boldsymbol{y}_{\rm model}(\boldsymbol{\theta})$.}---
Direct evaluation of \textsc{cujet} for arbitrary parameter set $(\boldsymbol{\theta})$ throughout posterior sampling is computationally prohibitive. We therefore estimate the model prediction, $\boldsymbol{y}_{\rm model}(\boldsymbol{\theta})$, by constructing a Gaussian-process emulator (GPE) using \textsc{gpytorch}~\cite{gardner2018gpytorch}. The training design contains 666 parameter points generated by maximin Latin-hypercube sampling in the 12-dimensional parameter space, together with 84 independent points reserved for validation.

At each design point, all $R_{\rm AA}$ and $v_2$ predictions entering the calibration are combined into a common output vector. The outputs are standardized separately for each observable, collision system, and centrality class, and are subsequently compressed through principal-component analysis. The first 15 principal components are retained, and an independent Gaussian process with an automatic-relevance-determination radial-basis-function kernel is trained for each principal-component coefficient. The emulator is trained directly on the \textsc{cujet} observables, without a logarithmic transformation or subtraction of a reference model.

The emulator is tested against full \textsc{cujet} calculations at the 84 independent validation points, corresponding to 8400 individual observable values. As shown in Fig.~\ref{fig:gpe_validation}, the emulator predictions closely follow the full-model results across all collision systems and observables. Over the complete validation set, the root-mean-square error is $2.43\times10^{-3}$, with aggregate values of $3.04\times10^{-3}$ for $R_{\rm AA}$ and $5.25\times10^{-4}$ for $v_2$, demonstrating good overall predictive accuracy.

The empirical coverage probabilities of the nominal $1\sigma$ and 95\% predictive intervals are 0.744 and 0.925, respectively. The standardized residuals have a mean of $-0.031$ and a standard deviation of 1.11. These results indicate negligible overall bias, while the predictive uncertainty is slightly underestimated. We account for this residual miscalibration through the additional $\sigma_{\rm noise}$ contribution in Eq.~\eqref{eq:model_uncertainty}.

\vspace{3mm}
\emph{Covariance matrix.} --- 
The covariance matrix encodes both experimental and theoretical uncertainties, either correlated or uncorrelated between different data points.
Because complete experimental covariance matrices are not available for all datasets, the baseline analysis assumes diagonal uncertainties,
\begin{equation}
\Sigma_{i}
=
\sigma_{{\rm exp},i}^{\,2}
+
\sigma_{{\rm model},i}^{\,2},
\label{eq:baseline_covariance}
\end{equation}
where the experimental statistical and systematic uncertainties are combined in quadrature. The model uncertainty is written as
\begin{equation}
\sigma_{{\rm model},i}^{\,2}
=
\sigma_{{\rm GPE},i}^{\,2}
+
\sigma_{\rm noise}^{\,2}.
\label{eq:model_uncertainty}
\end{equation}
Here, $\sigma_{{\rm GPE},i}$ is the predictive uncertainty of the Gaussian-process emulator, while $\sigma_{\rm noise}$ is a global empirical error floor estimated from the independent validation set in the block-wise standardized observable space. Since the model outputs are standardized separately for each observable, collision system, and centrality class before training the emulator, the validation residuals have a common dimensionless scale across the different observable blocks. The error floor is defined as
\begin{equation}
\sigma_{\rm noise}
=
\left(
\frac{1}{N_{\rm res}}
\sum_{j=1}^{N_{\rm res}}
\left(
\tilde y_{{\rm CUJET},j}
-
\tilde y_{{\rm GPE},j}
\right)^2
\right)^{1/2},
\end{equation}
where the tilde denotes the block-wise standardized observable and $N_{\rm res}$ is the total number of validation residuals. The resulting error floor is transformed back to the corresponding observable scale for each data point and added in quadrature to the GPE predictive uncertainty.

To assess the sensitivity of the inference to correlated experimental systematics, we additionally introduce a block-wise covariance model. For each dataset $k$, corresponding to a fixed collision system, centrality, and observable, we take
\begin{equation}
\boldsymbol{\Sigma}_k
=
\operatorname{diag}
 \left( \boldsymbol{\sigma}_{{\rm exp},k}^{\,2}
+ \boldsymbol{\sigma}_{{\rm model},k}^{\,2} \right)
+ \tau_k^2\; \boldsymbol{y}_{{\rm exp},k}
\boldsymbol{y}_{{\rm exp},k}^{\mathrm T},
\label{eq:correlated_covariance}
\end{equation}
where $\boldsymbol{y}_{{\rm exp},k}$ denotes the vector of experimental central values for dataset $k$. The rank-one term represents an assumed fully correlated relative normalization uncertainty within that dataset. Since complete experimental covariance matrices are unavailable, this construction is used only as a robustness test rather than as a detailed reconstruction of the experimental covariance structure.

We have examined the sensitivity of the posterior to the correlated-uncertainty model introduced in the previous section. Assuming coherent relative normalization uncertainties of $\tau_k=0.05$ and $0.10$ moderately broadens the posterior distributions, while leaving their peak locations and the inferred temperature dependence of $\chi_T(T)$ stable. See Figs.~\ref{fig:CoV_05} and~\ref{fig:CoV_10}. These values are not inferred experimental normalization uncertainties, but are introduced only to test sensitivity to plausible correlated systematics. We therefore adopt the baseline analysis with diagonal experimental uncertainties.

\vspace{5mm}
\emph{Posterior parameter distributions.} --- 
Figure~\ref{fig:posteriors} presents the marginal and pairwise posterior distributions obtained in the unconstrained and monotonic scenarios.

The coupling and screening parameters $\alpha_C$ and $c_M$ are well constrained and yield mutually consistent distributions in the two scenarios. Notably, the posterior coupling parameter is centered near $\alpha_C\simeq0.4$, considerably below the value $\alpha_C\simeq0.9$ obtained in earlier \textsc{cujet3.1} analyses employing fixed $\chi_T^u$ or $\chi_T^L$ parameterizations. The larger magnetic abundance inferred here provides additional jet--medium scattering strength and therefore reduces the value of the infrared coupling parameter required to describe the data.

In the unconstrained analysis, the low-temperature nodes of $\chi_T(T)$ remain clearly below unity and display an increasing trend with temperature. The data therefore favor a substantial chromo-magnetic fraction in the near-$T_c$ region without requiring monotonicity as an input.

At higher temperatures, the marginal distributions of $\chi_T(T)$ become progressively broader. This loss of sensitivity reflects the limited spacetime volume occupied by the highest-temperature regions of the fireball, even in the most central Pb--Pb collisions at $\sqrt{s_{NN}}=5.02~\mathrm{TeV}$. Consequently, the available jet-quenching measurements constrain the near-transition medium more strongly than the early, hottest stage.

Imposing the monotonic condition in Eq.~\eqref{eq:monotonic_prior} substantially reduces the high-temperature uncertainty while preserving the behavior preferred by the unconstrained posterior below approximately $0.35~\mathrm{GeV}$. The agreement between the two analyses in this data-sensitive region shows that the increasing trend of $\chi_T(T)$ is primarily driven by the measurements rather than imposed solely by the monotonic prior. We therefore adopt the monotonic posterior for the subsequent quantitative predictions.

\vspace{5mm}
\emph{Existence of chromo-magneto monopoles.} --- 
Based on the \textsc{cujet} model, our BA favored the presence of the CMMs. A comparison with the no-CMM baseline made this clear.
To determine whether the improved description associated with the magnetic component compensates for the additional model flexibility, we perform an explicit Bayesian comparison between two scenarios. The no-CMM baseline, denoted by $M_0$, is obtained by setting $\chi_T(T)=1$, for which the monopole density vanishes. In this limit, $c_M$ no longer affects the model predictions, and $\alpha_C$ is the only free parameter. The alternative model $M_1$ contains the monotonic temperature-dependent CMM fraction, together with the coupling and screening parameters $\alpha_C$ and $c_M$. For the no-CMM baseline, we obtained a posterior for $\alpha_C$ peaked at $1.07$, while the $68\%$ and $95\%$ credible intervals are $[1.00, 1.11]$ and $[0.92, 1.16]$, respectively.

\begin{figure}[t]
\centering
\includegraphics[width=\linewidth]{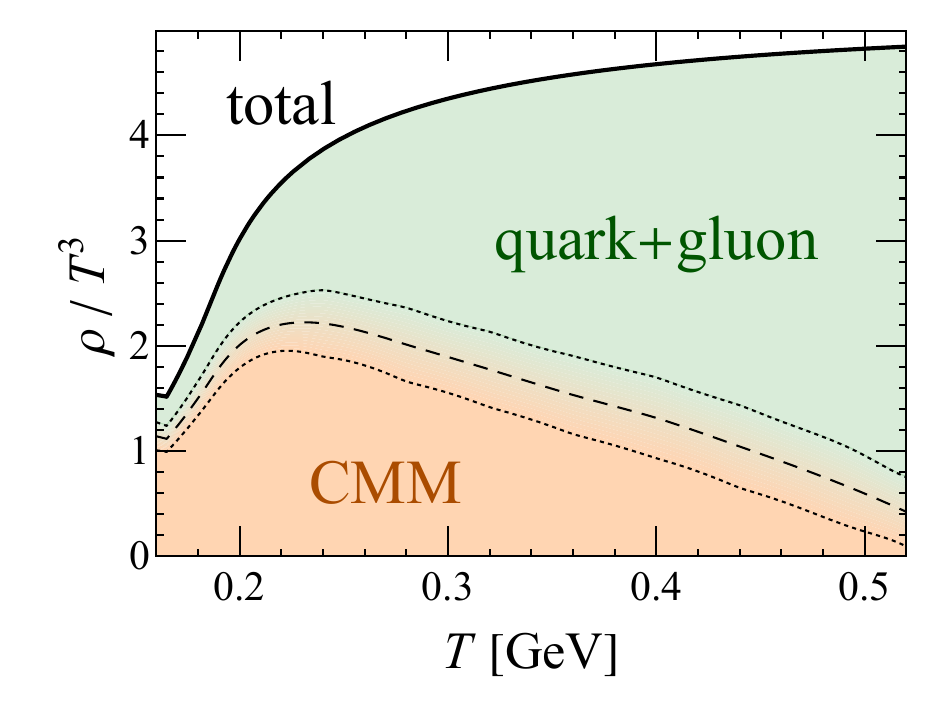}
\caption{Normalized densities of chromo-magnetic monopoles, chromo-electric quarks and gluons, and the total quasiparticle density from BA with monotonic constraint. Dotted lines indicate 68\% credible intervals.}
\label{fig:chiT_T}
\end{figure}
A statistically rigorous comparison between $M_1$ and $M_0$ should be drawn by calculating the Bayes factor.
The Bayesian evidence for model $M_i$ is
\begin{equation}
p(\mathcal D | M_i)
=
\int p(\mathcal D | \boldsymbol{\theta}_i,M_i)\, p(\boldsymbol{\theta}_i | M_i) \,\mathrm{d}\boldsymbol{\theta}_i ,
\label{eq:evidence}
\end{equation}
and the Bayes factor comparing the two models is
\begin{equation}
B_{10}
=
\frac{p(\mathcal D | M_1)}{p(\mathcal D | M_0)}.
\label{eq:bayes_factor}
\end{equation}

Using adaptive Monte Carlo integration with the same emulator-based likelihood as employed in the posterior inference, including the emulator uncertainty described above, we obtain
\begin{equation}
\ln B_{10}\simeq4.99,
\qquad
B_{10}\simeq1.47\times10^2.
\end{equation}
Since $B_{10}>1$, the data favor $M_1$, the data have a marginal likelihood about $147$ times larger under $M_1$ than under $M_0$. On the commonly used Jeffreys--Kass--Raftery scale, $\ln B_{10}\simeq4.99$ corresponds to strong evidence in favor of $M_1$. 
The Bayesian evidence therefore favors a temperature-dependent chromo-magnetic component in the \textsc{cujet} model, even after accounting for its larger parameter space. 

\vspace{5mm}
Imposing the monotonic posterior, the inferred temperature dependence of the chromo-magnetic and chromo-electric particle densities is shown in Fig.~\ref{fig:chiT_T}. The CMM density rises rapidly above $T_c$, reaches its maximum near $T\simeq1.5T_c$, and becomes the largest single contribution to the quasiparticle density over part of the near-transition region. At $T\gtrsim2T_c$, chromo-electric quarks and gluons gradually become dominant.

\begin{figure}[!hbtp]
\centering
\includegraphics[width=.45\textwidth]{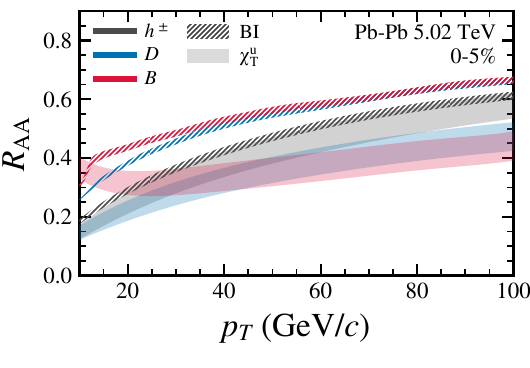}
\caption{
Nuclear modification factors of light hadrons (gray), $D$ mesons (blue), and $B$ mesons (red) in 0--5\% Pb--Pb collisions at $\sqrt{s_{NN}}=5.02~\mathrm{TeV}$. Hatched bands are the most recent BI results, solid opaque bands are for previous \textsc{cujet3} results with CMM fraction estimated by quark susceptibilities ($\chi_T^u$).
}
\label{fig:flavor_comparison}
\end{figure}
\section{Heavy-quark energy loss}
\label{sec:heavy_flavor}
\subsection{Jet Quenching Observables} 

\begin{figure}[!hbtp]
\centering
\includegraphics[width=.45\textwidth]{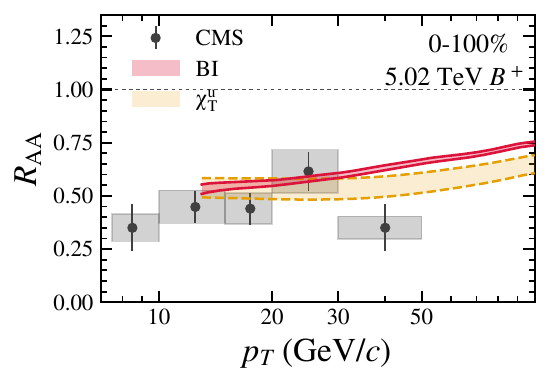}
\includegraphics[width=.45\textwidth]{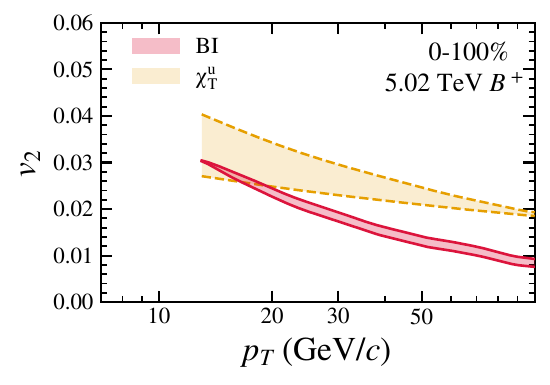}
\caption{
Predictions for $B^+$ mesons in minimum-bias Pb--Pb collisions at $\sqrt{s_{NN}}=5.02~\mathrm{TeV}$. The upper panel shows the nuclear modification factor $R_{\rm AA}$, with experimental data from CMS~\cite{CMS:2024vip}, while the lowe panel shows the corresponding prediction for $v_2$. The BI bands show the envelope of full \textsc{cujet} calculations obtained from the same representative parameter sets constructed from the marginal $1\sigma$ posterior bounds.
}
\label{fig:heavy_pbpb5020_bplus}
\end{figure}

\begin{figure}[!hbtp]
\centering
\includegraphics[width=.45\textwidth]{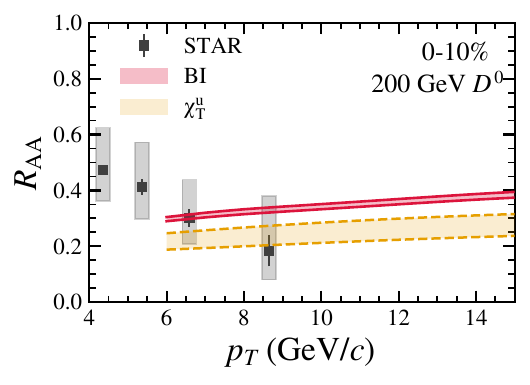}
\includegraphics[width=.45\textwidth]{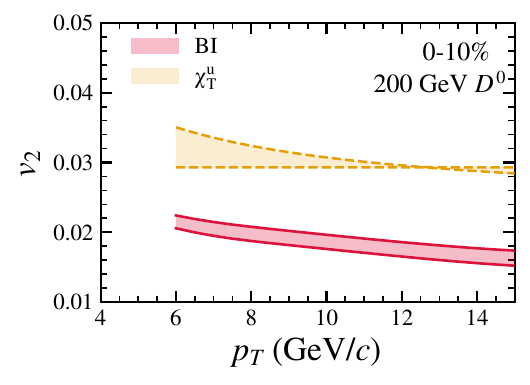}
\caption{
Predictions for $D^0$ mesons in 0--10\% Au--Au collisions at
$\sqrt{s_{NN}}=200~\mathrm{GeV}$. The upper panel shows the nuclear modification factor $R_{\rm AA}$, with experimental data from STAR~\cite{STAR:2018zdy}, while the lower panel shows the corresponding prediction for $v_2$. The BI bands show the envelope of full \textsc{cujet} calculations obtained from the same representative parameter sets constructed from the marginal $1\sigma$ posterior bounds.
}
\label{fig:heavy_auau200}
\end{figure}

With the CMM abundance and interaction parameters constrained by light-flavor charged hadrons, now we compute quenching of energetic heavy-flavor particles compared with corresponding experimental measurements. The same posterior-constrained values of $\chi_T(T)$, $\alpha_C$, and $c_M$ are used for charm- and bottom-quark calculations, without any additional calibration of the medium parameters.

Let us begin with a direct comparison of the $R_{\rm AA}$'s of light and heavy flavor particles. See Fig.~\ref{fig:flavor_comparison}. Results with the most updated BI parameters exhibit the flavor hierarchy, $R_{\rm AA}^{B} > R_{\rm AA}^{D} > R_{\rm AA}^{h^\pm}$, consistent with the dead-cone effect~\cite{Dokshitzer:2001zm, ALICE:2021aqk}, and the hierarchy gradually weakens toward high $p_T$. This hierarchy was not shown in previous \textsc{cujet3}-$\chi_T^u$ results with CMM fraction estimated by quark susceptibilities, see opaque bands. We owe the appearance of flavor hierarchy to the reduction of elastic scattering, as a consequence of the reduced $\alpha_C$.

\begin{figure*}[!hbtp]
\centering
\includegraphics[width=0.96\textwidth]{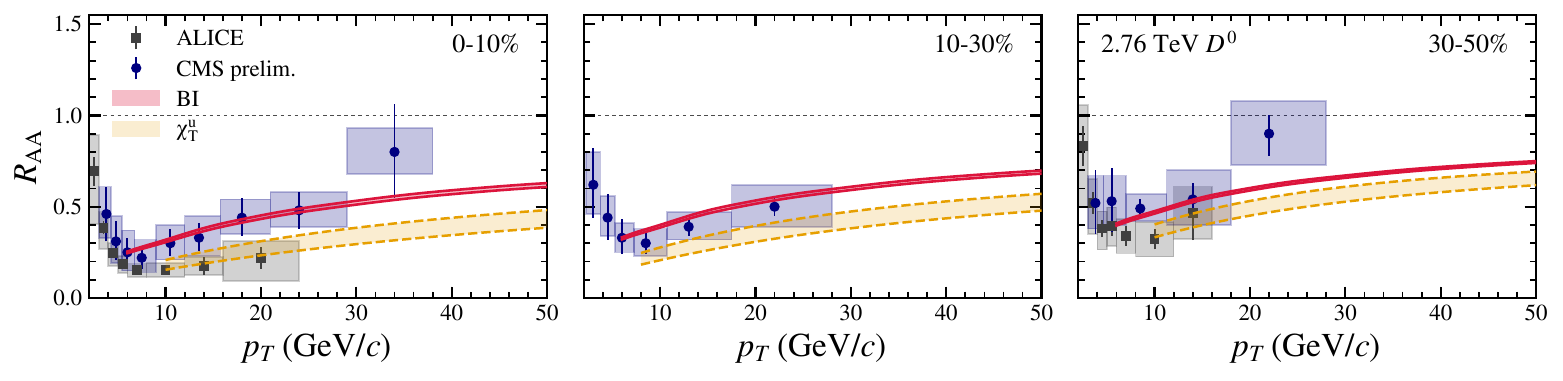}
\vspace{0.4em}
\includegraphics[width=0.96\textwidth]{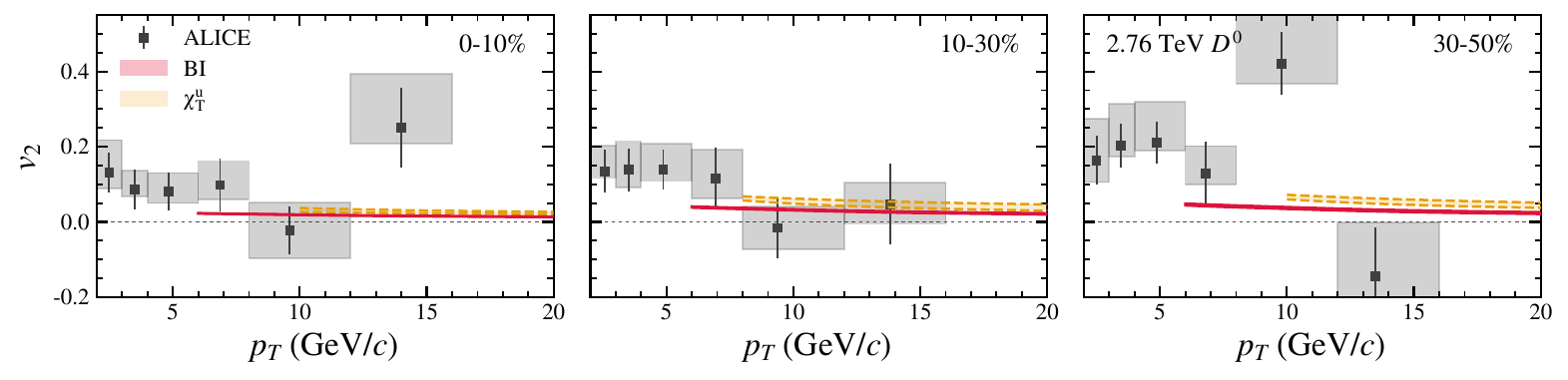}
\caption{
Nuclear modification factor $R_{\rm AA}$ (upper row) and elliptic flow coefficient $v_2$ (lower row) of $D^0$ mesons in Pb--Pb collisions at $\sqrt{s_{NN}}=2.76~\mathrm{TeV}$ for the 0--10\%, 10--30\%, and 30--50\% centrality intervals. The $R_{\rm AA}$ data are from ALICE~\cite{ALICE:2015vxz} and preliminary CMS measurements~\cite{CMS:2015hca}, while the $v_2$ data are from ALICE~\cite{ALICE:2014qvj}. The shaded region shows the envelope of full \textsc{cujet} calculations obtained from the same representative parameter sets constructed from the marginal $1\sigma$ posterior bounds.
\label{fig:heavy_pbpb2760}
}
\includegraphics[width=0.64\textwidth]{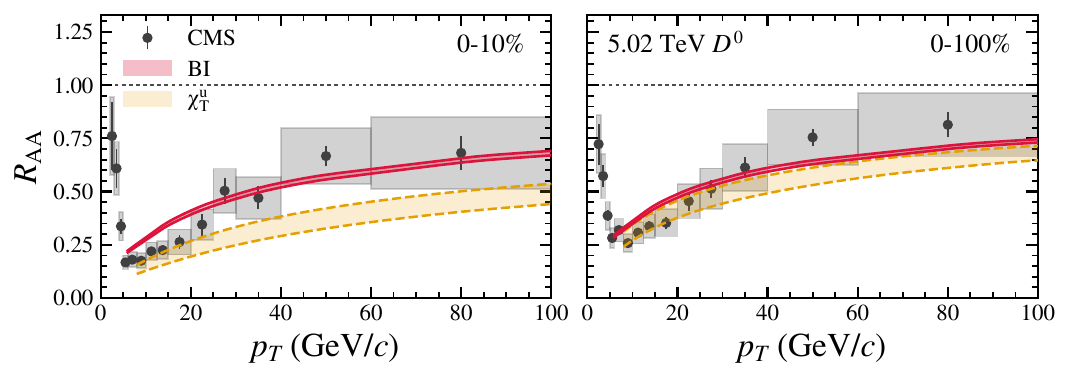}
\vspace{0.4em}
\includegraphics[width=0.64\textwidth]{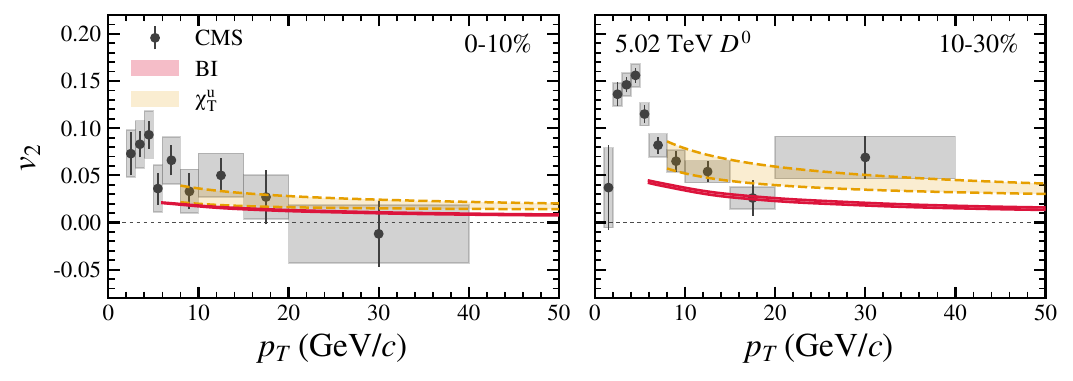}
\caption{
Nuclear modification factor $R_{\rm AA}$ (upper row) and elliptic flow coefficient $v_2$ (lower row) of $D^0$ mesons in Pb--Pb collisions at $\sqrt{s_{NN}}=5.02~\mathrm{TeV}$. The $R_{\rm AA}$ results are shown for the 0--10\% and 0--100\% centrality intervals, while the $v_2$ results are shown for the 0--10\% and 10--30\% intervals. The experimental measurements are from CMS~\cite{CMS:2017qjw,CMS:2017vhp}. The shaded region shows the envelope of full \textsc{cujet} calculations obtained from the same representative parameter sets constructed from the marginal $1\sigma$ posterior bounds.
\label{fig:heavy_pbpb5020_d0}
}
\end{figure*}

The bottom-flavor nuclear modification factor and elliptic flow are shown in Fig.~\ref{fig:heavy_pbpb5020_bplus}. The calculation predict substantial $B^+$-meson suppression and a gradual increase of $R_{\rm AA}$ with transverse momentum, with an overall magnitude compatible with the present experimental uncertainties. Results with the updated BI parameters exhibit less suppression of $B$ mesons, compared to the previous \textsc{cujet3}-$\chi_T^u$ parameter setting. The trend is consistent with the observation in Fig.~\ref{fig:flavor_comparison}.

Figure~\ref{fig:heavy_auau200} shows the $D^0$-meson $R_{\rm AA}$  and $v_2$ in central Au–Au collisions at $\sqrt{s_{NN}}=200~\mathrm{GeV}$. Like in the $B$ sector, results with the updated BI parameters exhibits higher $R_{\rm AA}^{D^0}$ than the previous \textsc{cujet3}-$\chi_T^u$ case, although both are broadly consistent with the available measurement in the high-$p_T$ region relevant to the present energy-loss calculation.

We further present the $D^0$-meson's $R_{\rm AA}$ and $v_2$ in Pb–Pb collisions with beam energy $\sqrt{s_{NN}}=2.76~\mathrm{TeV}$ and $5.02~\mathrm{TeV}$, respectively, in Figs.~\ref{fig:heavy_pbpb2760} and \ref{fig:heavy_pbpb5020_d0}. The calculations reproduce the overall suppression pattern and its centrality dependence.
The \textsc{cujet3}-BI parameters predicts lower $v_2$ than the \textsc{cujet3}-$\chi_T^u$ case, in accordance with the higher $R_{\rm AA}$ in the former case.

\subsection{Heavy Quark Diffusion Coefficient}
\label{sec:transport}
The Bayesian extraction of $\chi_T(T)$, $\alpha_C$, and $c_M$ provides a quantitatively constrained description of the chromo-electric and chromo-magnetic composition of the QGP. They are relevant to not only to the high-energy jet energy loss phenomenon but also to the low-energy transport coefficient, connected by the same microscopic scattering kernel.

\begin{figure}[!hbtp]
    \centering
    \includegraphics[width=0.45\textwidth]
    {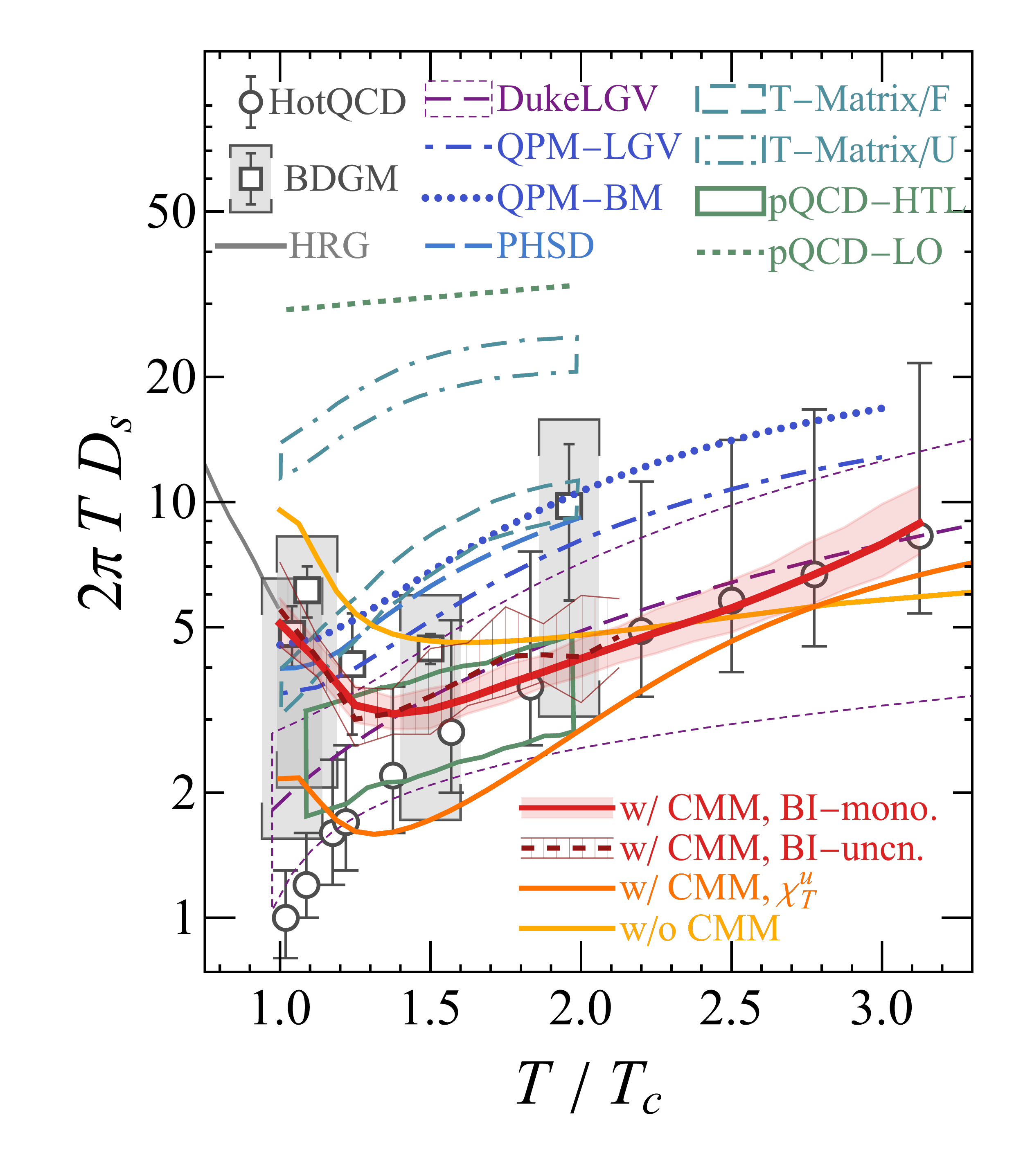}
    \caption{
    Temperature dependence of the dimensionless heavy-quark
    diffusion coefficient $2\pi T D_s$.
    Red curves and bands denote the posterior median and 95\% credible interval obtained in the present Bayesian analysis.
    Orange curves show the earlier \textsc{cujet3.1} results using
    the quark-susceptibility-based chromo-electric fraction
    $\chi_T^u$, while gold curves correspond to the
    no-CMM \textsc{cujet2} calculation.
    Results from lattice-QCD calculations
   ~\cite{Banerjee:2011ra,Ding:2012sp,HotQCD:2025fbd},
    a hadron-resonance-gas calculation~\cite{Lang:2012nqy},
    Langevin and quasiparticle approaches
   ~\cite{Cao:2011et,Cao:2012jt,Scardina:2017ipo},
    \textsc{phsd}
   ~\cite{Bratkovskaya:2011wp,Song:2015sfa,Song:2015ykw},
    $T$-matrix calculations
   ~\cite{vanHees:2007me,Riek:2010fk,Liu:2016ysz},
    and perturbative-QCD estimates
   ~\cite{Arnold:2003zc,Gossiaux:2008jv,Peshier:2008bg}
    are shown for comparison.
    }
    \label{fig:transport_coefficients}
\end{figure}

The heavy-quark spatial diffusion coefficient is estimated from
\begin{equation}
D_s(T)
=
\frac{4T^2}
{\hat q_F(E=3~\mathrm{GeV},T)},
\label{eq:Ds}
\end{equation}
where $\hat q_F(E,T)$ is the jet transport coefficient  that characterizes the average transverse-momentum broadening per unit path length for a propagating quark. Within the \textsc{cujet} framework, it is evaluated as
\begin{align}
\hat q_F(E,T)
={}&
\int_0^{6ET} \mathbf{q}_\perp^2\,
    \mathrm{d}^2\mathbf{q}_\perp
    \sum_f C_{f}\, \rho_f(T)\,
    \alpha_s^{\,i_f}(\mathbf{q}_\perp^2)
    \frac{\mathrm{d}\sigma_f}
    {\mathrm{d}\mathbf{q}_\perp^2},
\label{eq:qhat}
\end{align}
where $f\in\{g,q,m\}$. We take $i_f=0$ for scattering from chromo-electric quarks and gluons and $i_f=-2$ for scattering from CMMs. The color factor $C_{f}$ is $4/9$ for quark--quark scattering, and it takes unity for quark--gluon and quark--monopole channels.

The resulting $2\pi T D_s$ is shown in Fig.~\ref{fig:transport_coefficients}. 
The inferred $2\pi T D_s$ lies within the broad range obtained from lattice-QCD calculations and heavy-flavor transport models~\cite{Banerjee:2011ra,Ding:2012sp,HotQCD:2025fbd, Cao:2011et,Cao:2012jt,Scardina:2017ipo,Bratkovskaya:2011wp, Song:2015sfa,Song:2015ykw,vanHees:2007me,Riek:2010fk, Liu:2016ysz,Arnold:2003zc,Gossiaux:2008jv,Peshier:2008bg}. It develops a minimum close to the transition region, reflecting strong heavy-quark coupling to the medium. The smooth connection with the hadron-resonance-gas estimate below $T_c$~\cite{Lang:2012nqy} is suggestive of a continuous evolution from hadronic transport to strongly coupled partonic transport across the QCD crossover. Both these trends are robust against whether or not a monotonic constraint is taken for the temperature dependence of $\chi_T$.

The relation in Eq.~\eqref{eq:Ds} is a kinetic estimate based on an extrapolation of the \textsc{cujet} scattering kernel to thermal momentum scales. Their agreement with independent heavy-flavor analyses provides an important consistency test of the inferred medium. The resulting $2\pi T D_s$ is now ready to be implemented into a Boltzmann/Langevin simulation framework for heavy quarks---see e.g.~\cite{Li:2019wri, Li:2019lex}---so that a self-consistent prediction of heavy-flavor phenomenology in the low-to-intermediate $p_T$ region can be provided.

\section{Summary}
\label{sec:summary}

In this work, we have computed the heavy-flavor jet quenching observables by using the  \textsc{cujet3} framework, which has recently been improved by using Bayesian inference to extract the coupling parameters as well as the temperature-dependent chromo-electric and chromo-magnetic composition of the quark-gluon plasma~\cite{Guo:2025tsf} based on  the high-$p_T$ light-hadron $R_{\rm AA}$ and $v_2$ measurements from RHIC and the LHC. The Bayesian analysis~\cite{Guo:2025tsf} clearly suggests a substantial chromo-magnetic component in the near-transition region, with the inferred monopole density peaking around $T\sim1.5T_c$. 
Complementary details of the Bayesian analysis have been included in the present work, including verification for the prediction accuracy of the Gaussian process emulator and testing the robustness of our results against potential correlations among experimental data.

The main goal of this work is to utilize the heavy quark energy loss observables for an independent validation of the Bayesian-inference-improved CUJET3 model.  
Without any tuning  of parameters, the light-hadron-constrained posterior has been used to calculate the $D^0$- and $B^+$-meson observables. 
The results agree well with the data, where available.  
Compared to the previous \textsc{cujet3} model~\cite{Shi:2018izg, Shi:2018lsf} that  estimates the chromo-magneto monopole fraction with the quark susceptibility~\cite{McLerran:1987pz,Gottlieb:1988cq,Gavai:1989ce,Gottlieb:1987ac}, the current Bayesian-inference-improved model features a considerably smaller  coupling parameter $\alpha_C$ and consequently a reduced energy loss through the elastic scattering channel. This latter point is particularly relevant for heavy quarks and has allowed the results in the present work to exhibit the dead-cone effect, i.e., mass hierarchy in the nuclear modification factor, which was absent in the previous \textsc{cujet3} model. 

Finally, the same Bayesian-inference-improved CUJET3 model is used to estimate the heavy-quark spatial diffusion coefficient $2\pi T D_s$, which is an important QGP transport property and controls the in-medium evolution of low-$p_T$ heavy quarks. The predicted temperature-dependent diffusion coefficient is found to be broadly consistent with various phenomenological extractions and lattice-QCD constraints. 

Taken together, the phenomenological success of the current CUJET3 model in coherently describing both light- and heavy-flavor energy loss data strongly supports a microscopic picture of the QGP that features a significant component of chromo-magnetic monopoles in the $1\sim2T_c$ temperature region. 

\vspace{2mm}
\textbf{Acknowledgments.} --- JL and SS thank Dr. Miklos Gyulassy and Dr. Jiechen Xu for previous collaborations that developed the CUJET3 framework. The authors also thank Shuhan Zheng for helpful discussions.
YG and SS acknowledge support from the National Key Research and
Development Program of China under Contract No. 2024YFA1610700. JL is supported by the U.S. NSF under Grant No. PHY-2514992. SS also acknowledges supports from NSFC under grant No. 12575143 and Tsinghua University under Grant Nos. 04200500123, 531205006, and 533305009.

\bibliography{ref}

\clearpage

\begin{appendix}
\begin{widetext}
\section{Light-hadron observables}
\label{sec:light_hadron}

We show the light-hadron observables obtained with the medium composition constrained by the Bayesian analysis~\cite{Guo:2025tsf}. Representative parameter sets constructed from the marginal $1\sigma$ bounds of the posterior distributions are evaluated with the full \textsc{cujet} model to calculate the nuclear modification factor $R_{\rm AA}$ and the high-$p_T$ elliptic coefficient $v_2$ in Au--Au collisions at RHIC and Pb--Pb collisions at the LHC.
The Bayesian calibration includes the 0--10\% and 20--30\% centrality classes for Au--Au collisions at $\sqrt{s_{NN}}=200~\mathrm{GeV}$, the 0--10\% and 20--30\% classes for Pb--Pb collisions at $\sqrt{s_{NN}}=2.76~\mathrm{TeV}$, and the 0--5\% class for Pb--Pb collisions at $\sqrt{s_{NN}}=5.02~\mathrm{TeV}$. The remaining centrality intervals shown below were not used in the calibration and therefore provide additional predictions of the inferred medium. The shaded regions show the envelope of full \textsc{cujet} calculations obtained from these representative marginal-posterior parameter variations and are not intended as joint posterior credible intervals. For comparison, we also show the no-CMM baseline, corresponding to $\alpha_C = 1.07$ and $\chi_T(T)=0$, as well as the previous \textsc{cujet3} results that estimates the CMM fraction by quark number susceptibilities.

\begin{figure*}[!htbp]
\centering
\includegraphics[width=0.8\textwidth]{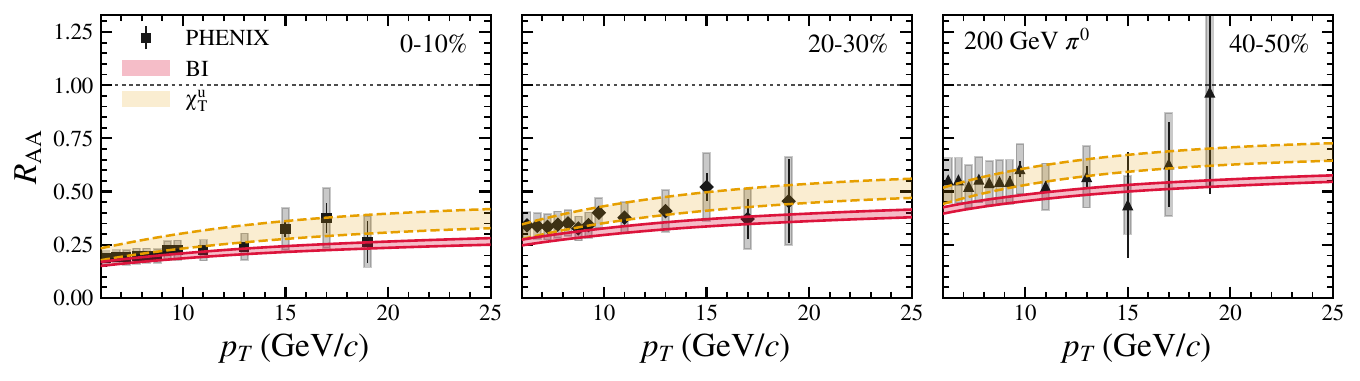}
\par\medskip
\includegraphics[width=0.8\textwidth]{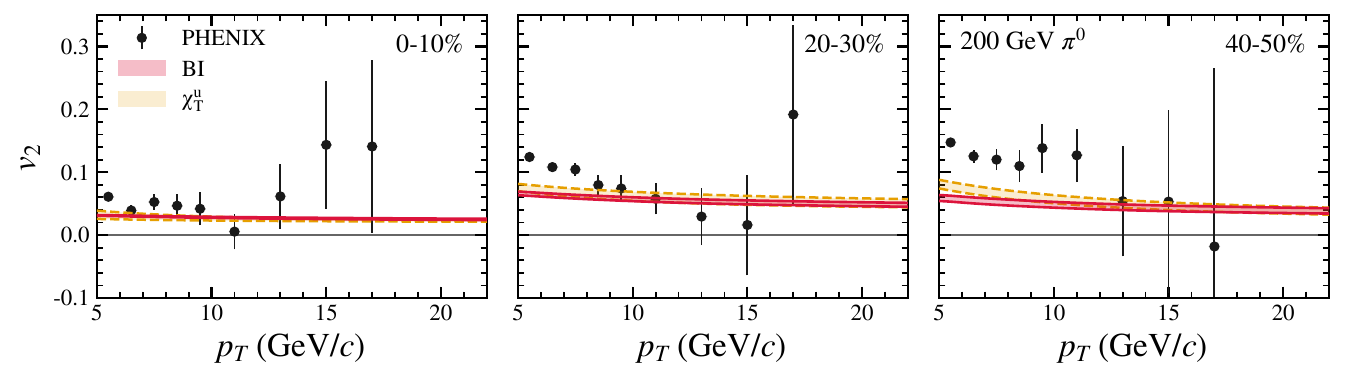}
\caption{
Neutral-pion nuclear modification factor $R_{\rm AA}$ (upper row) and elliptic flow coefficient $v_2$ (lower row) in Au--Au collisions at $\sqrt{s_{NN}}=200~\mathrm{GeV}$ for the 0--10\%, 20--30\%, and 40--50\% centrality intervals. The experimental measurements are from PHENIX~\cite{Adare:2008qa, Adare:2012wg, PHENIX:2013yhu}. The shaded regions show the envelope of full \textsc{cujet} calculations obtained from representative parameter sets constructed from the marginal $1\sigma$ posterior bounds. The 0--10\% and 20--30\% classes enter the Bayesian calibration, while the 40--50\% result is an additional prediction.
}
\label{fig:light_auau200}
\end{figure*}

\begin{figure*}[!hbpt]
\centering
\includegraphics[width=0.8\textwidth]{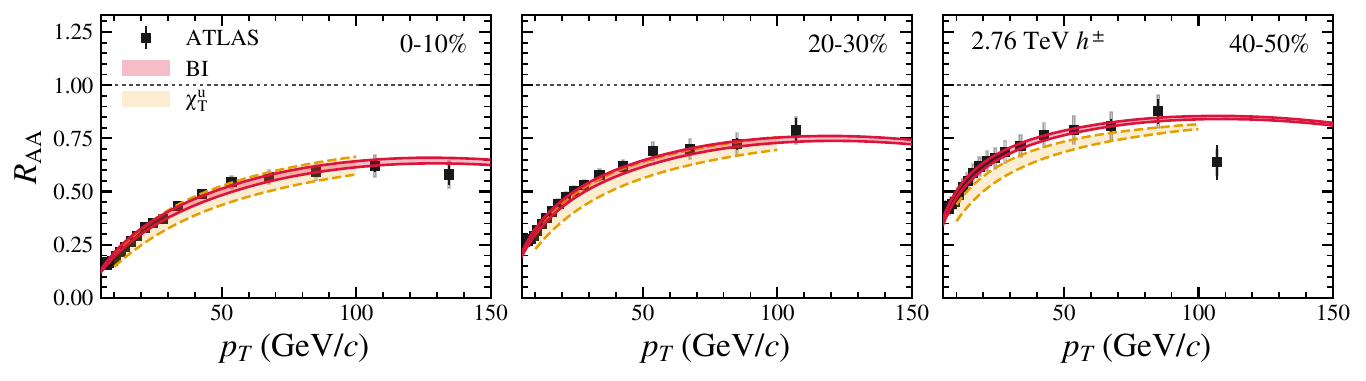}
\par\medskip
\includegraphics[width=0.8\textwidth]{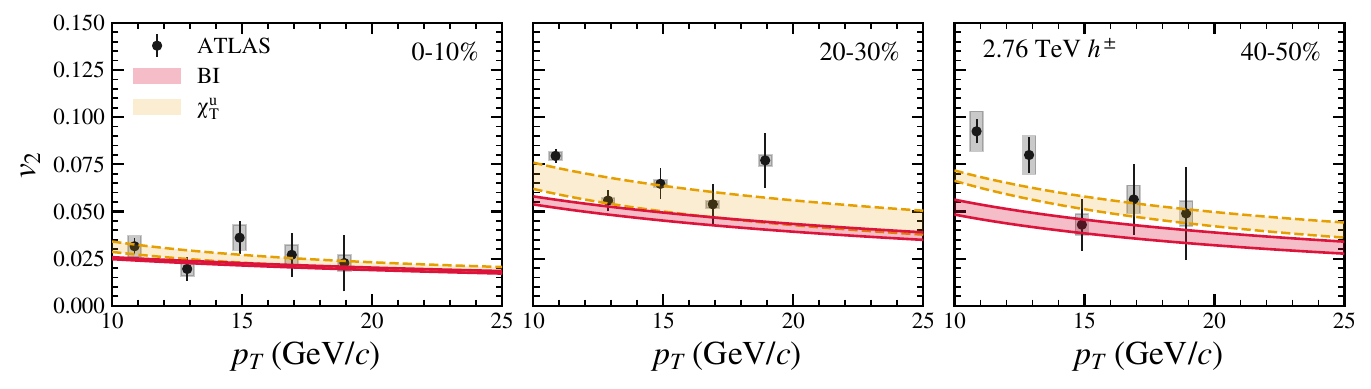}
\caption{
Same as \protect{Fig.~\ref{fig:light_auau200}} but for 0--10\%, 20--30\%, and 40--50\% most central Pb--Pb collisions at $\sqrt{s_{NN}}=2.76~\mathrm{TeV}$. The experimental measurements are from ATLAS~\cite{ATLAS:2015qmb, ATLAS:2011ah}.
}
\label{fig:light_pbpb2760}
\end{figure*}

\begin{figure*}[!hbpt]
\centering
\includegraphics[width=0.8\textwidth]{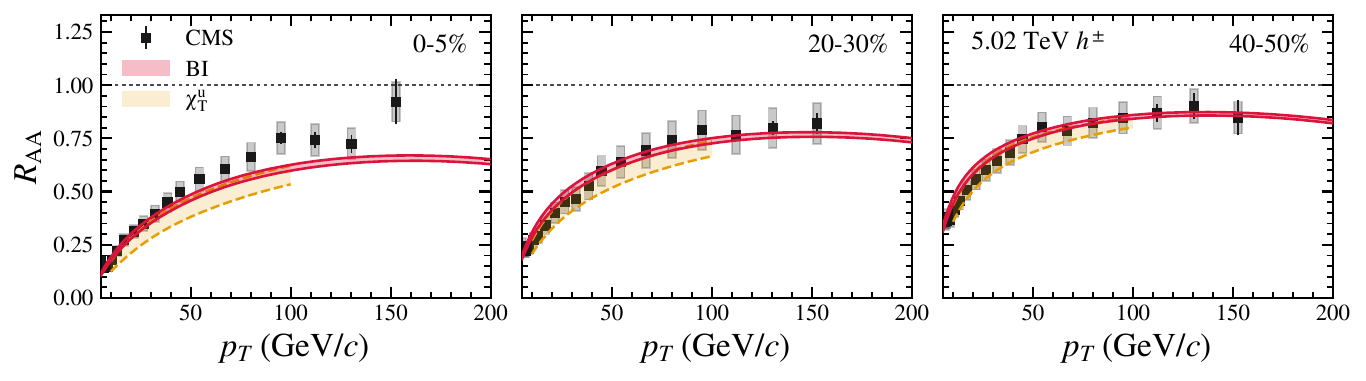}
\par\medskip
\includegraphics[width=0.8\textwidth]{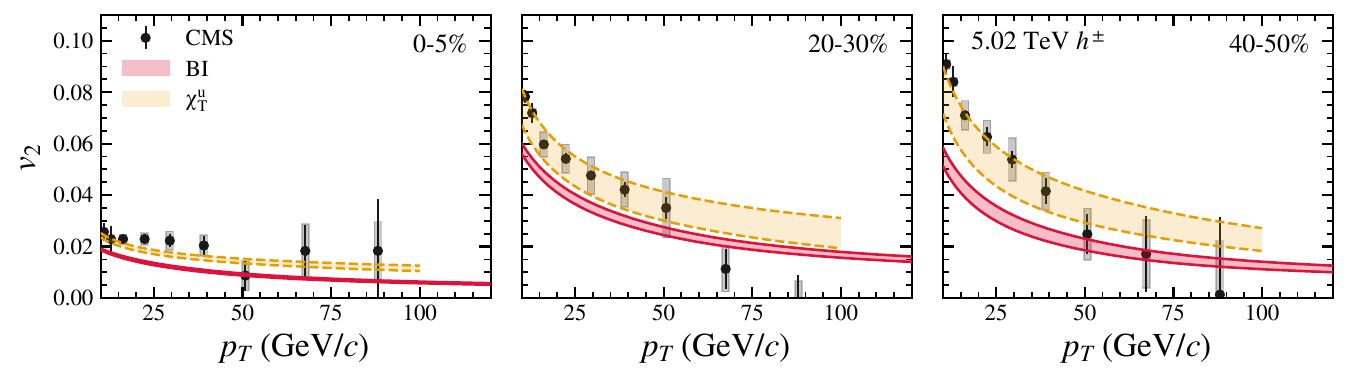}
\caption{
Same as \protect{Fig.~\ref{fig:light_auau200}} but for 0--10\%, 20--30\%, and 40--50\% most central Pb--Pb collisions at $\sqrt{s_{NN}}=5.02~\mathrm{TeV}$. The experimental measurements are from CMS~\cite{CMS:2012aa, CMS:2012zex, Khachatryan:2016odn, Sirunyan:2017pan}. Only the 0--5\% class enters the Bayesian calibration; the 20--30\% and 40--50\% results are additional predictions.
}
\label{fig:light_pbpb5020}
\end{figure*}

Figure~\ref{fig:light_auau200} shows the neutral-pion results at RHIC. The calculations reproduce the rise of $R_{\rm AA}$ with transverse momentum and the weakening of suppression from central to more peripheral collisions. The agreement extends to the 40--50\% centrality class, which was not included in the calibration. The calculated $v_2$ is broadly compatible with the PHENIX measurements, although the experimental uncertainties become large at the highest transverse momenta.

At $\sqrt{s_{NN}}=2.76~\mathrm{TeV}$, shown in Fig.~\ref{fig:light_pbpb2760}, the calculations describe the overall magnitude, momentum dependence, and centrality ordering of the charged-hadron suppression. The predicted 40--50\% result is also consistent with the observed reduction of suppression toward peripheral collisions. The main momentum and centrality dependence of $v_2$ is reproduced, although the calculation underestimates part of the noncentral data at moderate transverse momentum.

The corresponding results at $\sqrt{s_{NN}}=5.02~\mathrm{TeV}$ are shown in Fig.~\ref{fig:light_pbpb5020}. The calculations reproduce the main features of $R_{\rm AA}$ in both the calibrated 0--5\% class and the predicted 20--30\% and 40--50\% intervals. For $v_2$, the model captures the qualitative momentum and centrality dependence but underestimates part of the noncentral data at moderate transverse momentum. The agreement improves toward higher $p_T$, where the partonic energy-loss description is expected to be more reliable.

Taken together, these comparisons show that a common posterior-constrained medium provides a consistent description of the overall suppression strength across collision energies and centralities, including several centrality classes not used in the Bayesian calibration. The same medium also accounts for a substantial part of the measured high-$p_T$ anisotropy, although some noncentral $v_2$ data at moderate transverse momentum remain underestimated. Within the \textsc{cujet} framework, this simultaneous description is associated with enhanced jet--medium interactions near $T_c$ and the sizable chromo-magnetic component favored by the Bayesian analysis.
\end{widetext}
\end{appendix}

\end{document}